\documentclass[a4paper,11pt]{article}
\usepackage{url}
\usepackage{parskip}
\usepackage{amsmath,amsfonts,amssymb,theorem,verbatim}
\usepackage{amsfonts,amssymb,graphics,theorem}
\usepackage[pdftex]{graphicx}
\usepackage{here}
\usepackage[greek, english]{babel}

\usepackage{float,multirow}
\floatstyle{ruled}
\newfloat{algo}{thp}{lop}
\floatname{algo}{Algorithm}

\DeclareMathOperator{\R}{\mathbb{R}}
\DeclareMathOperator{\N}{\mathbb{N}}

\newtheorem{Cor}{Corollary} \newtheorem{Lemma}{Lemma}
\newtheorem{Prop}{Proposition} 
{\theorembodyfont{\normalfont}

}

\def\BEN{\begin{enumerate}}  \def\BI{\begin{itemize}}
\def\EEN{\end{enumerate}}   \def\EI{\end{itemize}}

\def\beq{\begin{eqnarray}} \def\eeq{\end{eqnarray}}

\def\al*#1{\begin{align*}#1\end{align*}}

\def\ga*#1{\begin{gather*}#1\end{gather*}}

\def\alat*#1#2{\begin{alignat*}{#1}#2\end{alignat*}}
\def\bea{\begin{eqnarray*}}
\def\eea{\end{eqnarray*}}
\def\ml*#1{\begin{multline*}#1\end{multline*}}

 \def\unl{\underline}

  \def\R{{\mathbb R}}
\def\N{{\mathbb N}}  

  \def\td{\text{\rm d}}

     \def\s{\sigma}

\newcommand{\exit}{{\mbox{\, \vspace{3mm}}} \hfill\mbox{$\square$}}

\begin{document}
\title{A method of moments approach to pricing double barrier
contracts driven by a general class of jump diffusions\footnote{
Research supported by EPSRC grant EP/D039053/1\newline The
research was largely carried out while the authors were based at
King's College London.
\newline {\it
Acknowledgement:} We thank Mihail Zervos for useful suggestions
and helpful conversations.
 }}
\author{Bjorn Eriksson\footnote{Corresponding author}\quad and \quad Martijn
Pistorius\\
Imperial College London,\\ Department of Mathematics,\\ South
Kensington Campus,\\ London SW7 2AZ, UK\\
Email: \{b.eriksson08,m.pistorius\}@imperial.ac.uk}
\date{}

\maketitle
\begin{abstract}
We present the method of moments approach to pricing barrier-type
options when the underlying is modelled by a general class of jump
diffusions. By general principles the option prices are linked to
certain infinite dimensional linear programming problems.
Subsequently approximating those systems by finite dimensional
linear programming problems, upper and lower bounds for the prices
of such options are found. As numerical illustration we apply the
method to the valuation of several barrier-type options (double
barrier knockout option, American corridor and double no touch)
under a number of different models, including a case with
deterministic interest rates, and compare with Monte Carlo
simulation results. In all cases we find tight bounds with short
execution times. Theoretical convergence results are also
provided.
\end{abstract}

\newpage
\section{Introduction}
Barrier and barrier-type options are among the most widely and
frequently traded exotic options, especially in the area of
Foreign Exchange, which makes their valuation an important topic.
For example, a double barrier option is cancelled depending on
whether or not two levels have been crossed before maturity. Since
the pay-off of a barrier option depends on the entire path of the
underlying, it is clear that its valuation is more involved than
that of a standard European type option.

A well-documented empirical observation is that financial returns
data typically possess features such as asymmetry, heavy tails and
excess kurtosis, which cannot be captured by the classical
geometric Brownian motion model (GBM). Related is the well known
fact that under the GBM model it is not possible to calibrate
option prices to the volatility surface. One of the successful
modifications that has been proposed is to introduce jumps in the
evolution and work with L\'{e}vy models. Popular examples of such
L\'{e}vy models are VG, CGMY, NIG, GH and KoBoL. This approach is
classical by now and we refer to the standard references
\cite{Schoutens}, \cite{ContTankov} and \cite{BoyLev} for further
financial motivations for the use of jump models, background and
references. In a separate development (see e.g. \cite{Crosby} and
\cite{Geman}) it was noted that commodity prices often display
features such as mean-reversion and jumps that are clearly not
captured by the geometric Brownian motion model, and it was
proposed to employ jump-diffusion models to incorporate those
effects.

The valuation of barrier options has attracted a good deal of
attention and there exists currently a body of literature dealing
with different aspects of pricing barrier options. In particular,
for double barrier options, \cite{GeYor} and \cite{Pelsser}
developed a Laplace transform approach in the geometric Brownian
motion setting. \cite{Sepp} derived semi-analytical expressions in
a jump-diffusion setting with exponential jumps, also using a
transform approach. \cite{CarrCrosby} considered double no touches
in a setting with exponential jumps, allowing the process dynamics
to change after a barrier is breached. \cite{Linetsky} used
eigenfunction expansions to price double barrier options in a CEV
setting.

The mentioned papers exploit specific features of the model under
consideration and can therefore not be readily generalized and
applied to a different settings. A general approach, based on a
characterization of the moments of the underlying process, was
followed by \cite{Lasser} to price a class of exotic options. In a
diffusion setting \cite{Lasser} derived upper and lower bounds for
the price of exotic options in terms of semidefinite programs, and
provided theoretical and numerical convergence results for these
bounds. Before that, using linear programming, \cite{Stockbridge}
developed a method of moments algorithm to calculate first exit
time probabilities and moments of a diffusion.

In this paper we will follow a method of moments approach to price
double barrier options in a general setting of a polynomial-type
jump-diffusion. We will also allow the rate of discounting, which
is typically taken to be constant, to be a function of time and
underlying. We will now briefly describe the method of moments
approach. The first step is to express the option price as an
integral with respect to two measures, the (discounted) expected
exit measure and the (discounted) expected occupation measure. The
former describes the law of the underlying at expiration or at
crossing the barrier, while the latter described the law of the
process until this moment. Restricting ourselves to pay-offs that
are piecewise polynomial functions of the underlying, the value
can then be expressed as a linear combination of moments of these
two measures. The moments of those two measures are subsequently
shown to satisfy an infinite dimensional linear system. To the
price can thus be associated the two linear programming problems
of minimization and maximization of the latter criterion over the
spaces of measures. By adding conditions on the moments that
guarantee that a given sequence is equal to the moments of a
measure, one is led to an infinite dimensional linear programming
problem or a semi-definite programming problem. By restricting to
a finite number of moments we arrive at a finite dimensional
linear programming problem or a semi-definite programming problem.

We will numerically illustrate this method for a American corridor
and double no touch and double knock-out option under different
models, by solving linear programming problems. In all cases we
find tight bounds, with short execution times. We also provide a
convergence proof to show that the values of the linear
programming problems converge monotonically to the value of the
option if the number of moments employed is increased.

The remainder of the paper is organized as follows. In Section
\ref{sec:prob} we specify the model and the problem setting.
Section \ref{sec:mm} is devoted to the method of moments,
describes the algorithm and provides a convergence proof.
Section \ref{sec:numerics} provides the implementation and numerical
examples. Proofs are deferred to the Appendix.

\section{Problem setting}\label{sec:prob}
Assume that the underlying $X$ evolves according to the SDE
\begin{align}\label{X}
  d X_t &= b(t,X_t)dt + \sigma(t,X_t)dW_t +
  \lambda(t,X_t)\td J_t,\quad  X_0 = x_0,
\end{align}
where $W$ is a Brownian motion and $J$ is a pure jump L\'{e}vy
process (that is, a process with independent stationary increments
without Gaussian component), and $\sigma$,$b$, and $\lambda$ are
given functions that will be specified below. In this setting we
will value a barrier option of knock-out type with pay-off $h(T,
X_T)$ at the maturity time $T$ if the underlying has not left a
set $B$ before time $T$ and that pays a stream of payments
$g(s,X_s)$ until the first moment $\tau_B$ that $X$ leaves $B$ or
time $T$, whichever comes earlier. Modelling the risk neutral
discounting as a function $r=r(t,X_t)$ of $t$ and $X_t$ it follows
by standard arbitrage pricing principles that the value $v$ of this
contract is given by
\begin{equation}\label{eq:value}
v=E\left[
e^{-\alpha_{T}}h(T,X_T)I_{\{T<\tau_B\}}+\int_0^{\tau_B\wedge T}
e^{-\alpha_s}g(s,X_s)ds\right],
\end{equation}
where $\tau_B=\inf\{t\ge0: X_t\notin B\}$ and
\begin{equation}\label{alpha}
\alpha_t = \int_0^t r(s,X_s)\td s.
\end{equation}
We will restrict ourselves to the case that the functions $h,g$
are piecewise polynomial functions, that is, for some partitions
$\{C_i\}$ and $\{D_i\}$ of $[0,T]\times\R$,
\begin{align}\label{gh}
  h(t,x) &= \sum_{i=1}^k h_i(t,x) I_{\{(t,x)\in C_i\}},\quad
g(t,x) = \sum_{i=1}^l g_i(t,x) I_{\{(t,x)\in D_i\}},
\end{align}
where $h_i, g_i$ are polynomials in $(t,x)$. Note that many
contracts have a pay-off function that is of this form, including
call and put options and straddles. We observe that we will then
be able to express the value $v$ in terms of moments of certain
probability measures, reducing the calculation of $v$ to the
calculation of these moments. Further, we will assume that $X$ is
a `polynomial' process, that is, in eqs. \eqref{X} and
\eqref{alpha}
$$
\text{$b(t,x)$, $\sigma^2(t,x)$, $\lambda(t,x)$ and  $r(t,x)$ are
polynomials,}
$$
such that \eqref{X} admits a unique (weak) solution. Associated to
$X$ is the infinitesimal generator that acts on functions $f$ in
its domain as
\begin{equation}\label{A}
Af = \frac{\partial f}{\partial t} + b\, \frac{\partial
f}{\partial x} + \frac{\sigma^2}{2}\, \frac{\partial^2 f}{\partial
x^2} + Bf,
\end{equation}
where $Bf$ is an integro-differential operator given by
\begin{equation}\label{B}
Bf(t,x) = \int_{\R} \left[f(x+\lambda(t,x)y) - f(t,x) -
\lambda(t,x)\frac{\partial f}{\partial x}(t,x)
yI_{\{|y|<1\}}\right]\Lambda(\td y),
\end{equation}
where $\Lambda$ denotes the L\'{e}vy jump measure of $X$.
Note that the
operator $A$ maps polynomials to polynomials, which is an
essential property needed in the moment approach, as shown in the
next section.

We next present some models that are included in our setting.

\bigskip

\noindent{\bf Examples.}
\begin{itemize}
\item The classical geometric Brownian motion satisfies the SDE
\begin{equation*}
dX=bXdt+\sigma XdW
\end{equation*}
where $W$ denotes a one-dimensional Brownian motion,
and has the infinitesimal generator
\begin{equation}\label{eq:genBGM}
Af(x) = bx\frac{df}{dx}+\frac{x^2\sigma^2}{2}\frac{d^2f}{dx^2}.
\end{equation}
\item L\'{e}vy models (for an overview see e.g. \cite{Schoutens}
or \cite{ContTankov}) For example, the variance gamma process
(\cite{MCC}) evolves according to $dX=b_1dt+dZ$, where $b_1$ is a
constant and $Z$ is a L\'{e}vy process with L\'evy measure
\begin{equation}\label{eq:VGeta}
\eta(dx)=\frac{C}{x}e^{-Mx}I_{\{x>0\}}dx +
\frac{C}{|x|}e^{-G|x|}I_{\{x<0\}}dx,
\end{equation}
with $C,G,M$ positive constants, and its infinitesimal
generator is specified by
\begin{align}\label{eq:genVG}
Af(x)=b\frac{df(x)}{dx}+\int\left[ f(x+y)-f(x)\right]\eta(dx).
\end{align}
where $b = b_1 - \int_{-1}^1 x\eta(dx)$.

\item Additive processes with polynomial time-dependent
coefficients (see e.g. \cite{ContTankov} for background), obtained
by taking $b$, $\s^2$ and $\lambda$ to be polynomials of $t$ only.
For example, $X$ evolving according to $dX = b_1(t)dt + dZ$ for a
L\'{e}vy process $Z$ and a polynomial $b_1(t)$.

\item Affine processes, obtained by taking $\lambda$ constant and
$b$, $\s^2$ affine functions in $x$, independent of $t$ (see e.g.
\cite{Cuchiero} for applications of affine models in finance). An
example of an affine diffusion is the Cox Ingersoll Ross (CIR)
model, which is a mean-reverting diffusion satisfying the SDE
\begin{align}\label{eq:SDECIR}
dX=a(b-X)dt+\sigma\sqrt{X}dW,\qquad a,b>0,
\end{align}
with the infinitesimal generator
\begin{align}\label{eq:genCIR}
Af(x)=a(b-x)\frac{d f}{d x}+ \frac{\sigma^2x}{2}\frac{d^2 f}{d
x^2}.
\end{align}
\end{itemize}

\section{Method of moments}\label{sec:mm}
Denoting by $\nu$ and $\mu$ the {\sl discounted exit location
measure} and the {\sl discounted occupation measure} given by
\begin{align*}
\nu(A) &= E[e^{-\alpha_\tau}I_{\{(\tau,X_\tau)\in A\}}], \quad
\mu(A) = E\left[\int_0^\tau e^{-\alpha_s}I_{\{(s,X_s)\in
A\}}ds\right],
\end{align*}
for Borel sets $A\in \mathcal B([0,T]\times\R)$, the value $v$ of
the contract can be expressed as
\begin{equation}\label{v}
v = \int h(t,x) \nu(dt,dx) + \int g(t,x)\mu(dt,dx)
\end{equation}
The measure $\mu$ describes the distribution of the process
$(t,X_t)$ before the stopping time $\tau$ whereas the measure
$\nu$ describes the distribution upon termination at $\tau$. For
example, in the case of a up-and-out barrier option at level
$B_u$, termination occurs if the barrier $B_u$ is crossed or the
maturity $T$ is reached.

In view of the form \eqref{gh} of $g$ and $h$, $v$ can be expressed
in terms of the moments of $\mu$ and $\nu$, as follows:
\begin{equation}\label{eq:Lmunu}
v = \sum_{i,j}\sum_m d_{i,j}(m) \nu_{i,j}^{(m)} + \sum_{i,j}\sum_m
b_{i,j}(m) \mu_{i,j}^{(m)},
\end{equation}%
where we denote by $m_{i,j}=\int t^ix^jm(dt,dx)$ the $ij$th moment
of a measure $m$ and by $\nu^{(m)}=\nu(\cdot\cap C_m)$ and
$\mu^{(m)}=\mu(\cdot\cap D_m)$ the restrictions of $\nu$ and $\mu$
to $C_m$ and $D_m$, and where $d_{i,j}(m)$ and $b_{i,j}(m)$ are
some constants.
%

\subsection{The adjoint equation}\label{sec:adjoint}
The measures $\mu$ and $\nu$ are closely related to each other and
to the generator of the underlying process $X$. Informally, for
suitably regular $f$ and all bounded stopping times $\tau$
Dynkin's lemma yields that
\begin{equation}\label{adjoint}
  E[e^{-\alpha_\tau}f(\tau, X_\tau)] - E[f(0,X_0)] =
  E\left[\int_0^\tau e^{-\alpha_t}(Af - r f)(t,X_t)dt\right],
\end{equation}
where $Af$ is given in \eqref{A}, which can be expressed in terms
of the measures $\nu$ and $\mu$ as
\begin{equation}\label{eq:intbae}
  \int f(t,x) \nu(dt,dx) = f(0,x_0) + \int (Af - r f)(t,x)\mu(dt,dx).
\end{equation}
The identity \eqref{eq:intbae} is called the \emph{basic adjoint
equation} (See e.g. \cite{Stockbridge}). As noted before, a formal
application of the generator shows that $A$ maps polynomials to
polynomials. More specifically, by applying \eqref{eq:intbae} to a
monomial $f_{ij}(t,x)=t^ix^j$ we obtain the following infinite
system of equations linking the moments of $\mu$ and $\nu$:
\begin{align}\label{eq:inbae}
\int t^ix^j\nu(dt,dx)- x_0^j1_{i=0} = \sum_{k,l}c_{k,l}(i,j)\int
t^kx^l\mu(dt,dx).
\end{align}
where $(Af_{ij}-rf_{ij})(t,x)=\sum_{k,l}c_{k,l}(i,j)t^lx^k$, or
equivalently, in compact notation,
\begin{align}\label{eq:bae}
\nu_{i,j}-x_0^j1_{i=0}=\sum_{k,l}c_{k,l}(i,j)\mu_{k,l}.
\end{align}
The following result provides sufficient conditions to justify
this informal analysis:
\begin{Prop}\label{prop:dynkin}
Suppose that for all $k=0,1,2,\ldots$
\begin{equation}\label{eq:int}
  \int |y|^k(1\wedge y^2)\Lambda(\td y) +
  E\left[\int_0^\tau e^{-\alpha_t}|X_t|^k\td t\right] < \infty.
\end{equation}
Then eqn. \eqref{eq:bae} holds for all $i,j=0,1,2,\ldots$.
\end{Prop}
The proof is deferred to the Appendix.

\medskip

{\bf Remark.} Partial barrier or forward starting barrier options
can also be included in this setting by slightly adapting the
definitions. With $\tilde\tau=\inf\{t\in[T_0,T]: X_t\notin B\}$
it holds that \begin{equation*}
  E[e^{-\alpha_{\tilde\tau}}f(\tilde\tau, X_{\tilde\tau})] - E[f(T_0,X_{T_0})] =
  E\left[\int_{T_0}^{\tilde\tau} e^{-\alpha_t}(Af - \alpha f)(t,X_t)dt\right]
\end{equation*}
which leads to the adjoint equation
\begin{equation*}
  \int f d\tilde\nu = \int f d\tilde\nu_0 + \int (Af - \alpha f)d\tilde\mu.
\end{equation*}
\subsubsection{Truncation}
We restrict ourselves now to contracts that are knocked out if $X$
leaves a finite interval, so that $v$ is given by \eqref{eq:value}
with $h(t,x)=0$ for $x\notin B:= [b_-, b_+]$. If the minimum
$\unl\lambda$ of $\lambda(t,x)$ over $[0,T]\times [b_-, b_+]$ is
strictly positive, it is possible to derive for such double knock
out contracts a modification of the adjoint equations that is
valid without integrability restrictions. To that end, note that a
double knock-out option becomes worthless at the first time that a
jump occurs of size larger than $L_+:=(b_+-b_-)/\unl\lambda$ or
smaller than $L_-:=(b_--b_+)/\unl\lambda$, since any such jump
will take $X$ out of the interval $[b_-,b_+]$. As such jumps occur
at a rate $\lambda_* = \Lambda(\R\backslash[L_-,L_+])$,
independent of the smaller size jumps and the diffusion part, it
follows that the value $v$ of the contract does not change if we
replace $\Lambda$ and $r$ by
$$
\tilde\Lambda = \Lambda(\cdot\cap[L_-,L_+]),\qquad\quad \tilde r =
r + \lambda_*,
$$
which corresponds to replacing the underlying $X$ by the process
$\tilde X$ that is `killed' when the first jump occurs with size
larger than $L_+$ or smaller than $L_-$. In summary, using
$\nu^{ij}$, $\mu^{ij}$ to denote the $ij$th moments of the exit
and occupation measures $\nu$, $\mu$ of the killed process $\tilde
X$, we have the following result (with a proof in the Appendix):
\begin{Cor}\label{cor:dynkin}
If $B=[b_-, b_+]$ and $\unl\lambda>0$, then $v$ is given by
\eqref{v} where $\nu^{ij}$ and $\mu^{ij}$ solve the system of
equations
\begin{equation}\label{eq:adftilde}
  \nu^{ij} - x_0^j1_{i=0} = \sum_{k,l} \tilde c_{k,l}(i,j) \mu_{k,l}
\end{equation}
where the coefficients $\tilde c_{k,l}(i,j)$ are defined by
$$\tilde Af_{ij} - \tilde r f_{ij} = \sum_{k,l}\tilde c_{k,l}(i,j)
t^kx^l,$$ with $\tilde A$ defined in \eqref{A}--\eqref{B} with
$\Lambda$ replaced by $\tilde\Lambda$.
\end{Cor}

\subsection{Linear programs}
By optimizing over the pair of measures that satisfies the adjoint
equations, the value $v$ can be bounded, as follows:
\begin{equation}\label{eq:LPI}
  \inf_{\nu,\mu} L(\nu,\mu) \leq v \leq \sup_{\nu,\mu} L(\nu,\mu)
\end{equation}
which concerns linear programs over the measures, since $L$ is the
linear functional of the moments of $\nu$ and $\mu$ given by
$$L(\nu,\mu):=\sum_{i,j}\sum_m d_{i,j}(m) \nu_{i,j}^{(m)} +
\sum_{i,j}\sum_m b_{i,j}(m) \mu_{i,j}^{(m)}$$ and the infimum and
the supremum are taken over the pairs of measures $(\nu,\mu)$
supported on $([0,T]\times \R\backslash B, [0,T]\times B)$ that
satisfy the linear adjoint equations derived before. To formulate
these optimization problems completely in terms of moment
sequences, we need to express the condition that $\mu$ and $\nu$
be measures in terms of their moments, as in general there is no
guarantee that any solution of the system \eqref{eq:bae} is the
moment sequence of some measure. The problem to determine whether
a given sequence is the moment sequence of some measure and, if
so, whether this measure is uniquely determined (in which case the
measure is called {\sl moment-determinate}) has been extensively
studied. It is known, see \cite{Stoy}, that the Cram\'er condition
\begin{align*}
\int_{\R} e^{c|x|}m(dx)<+\infty, \textrm{ for some } c>0
\end{align*}
is a sufficient condition for a measure $m$ to be moment
determinate. In particular, any measure with compact support is
moment-determinate. Further, the following Hausdorff conditions
are necessary and sufficient for a given sequence $m_i$ to
correspond to a moments of a measure $m$ with support on the
interval $[a,b]$ (see e.g. \cite{Feller}):
\begin{equation}\label{eq:H1}
\sum_{j=0}^n\binom{n}{j}(-1)^j \tilde m_{j+k}\geq 0 \quad \forall
n,k=0,1,2,...
\end{equation}
where the $\tilde m_i$ are linear combinations of the $m_j$, as
follows:
$$
\tilde m_l = (b-a)^{-l}\sum_{i=0}^l \binom{l}{i} (-a)^{l-i} m_i.
$$
In fact, the $\tilde m_i$ are themselves the moments of a measure
$\tilde m$ that is the affine transformation of $m$ supported on
$[0,1]$. That these conditions are necessary immediately follows
by observing that $\int_0^1y^k(1-y)^n\tilde m(dy)$ is non-negative
and by expressing $\tilde m_i$ in terms of $m_j$. More generally,
given an array $(m_{ij}, i,j=0,1,\ldots)$, the two dimensional
Hausdorff-conditions (e.g. \cite{ShohatTamarkin})
\begin{equation}\label{eq:H2}
\sum_{i=0}^m\sum_{j=0}^n\binom{m}{i}\binom{n}{j}(-1)^{i+j}\tilde
m_{i+l,j+k}\geq 0. \quad \forall n,m,k,l=0,1,2,...
\end{equation}
where the $\tilde m_{i,j}$ are related to the $m_{i,j}$ by
$$
\tilde m_{k,l} = (b-a)^{-k}(d-c)^{-l} \sum_{i=0}^k\sum_{j=0}^l
\binom{k}{i}\binom{l}{j}(-a)^{k-i}(-c)^{l-j} m_{i,j},
$$
are necessary and sufficient conditions to guarantee that there
exists a measure $m$ supported on $[a,b]\times[c,d]$ such that
$m_{ij} = \int_a^b\int_c^d x^iy^j m(dx,dy)$. See
\cite{ShohatTamarkin} for proofs and further background on
problems of moments.

\subsubsection{Unbounded support}\label{sec:qmc}
In the case that the measure has unbounded support there also
exists conditions to characterize a sequence of moments. These
conditions are no longer linear but can be conveniently be
formulated in terms of so-called {\sl moment} and {\sl localizing}
matrices (their definitions are recalled in the appendix). For a
sequence to be equal to the moments of some measure it is
necessary and sufficient that these matrices are positive
definite. See \cite{KreinNudel} or \cite{CurtoFial} for a proof of
this fact.

\subsection{Approximations and convergence}
To be able to calculate lower and upper bounds for the value $v$
we approximate the optimization problems in \eqref{eq:LPI} by
restricting the total number of moments used to $N$. If $B$ is a
finite interval, employing the moment conditions \eqref{eq:H1} and
\eqref{eq:H2} results in the following (finite) linear programming
problems:
\begin{equation*}
v^{(N)}_\pm := \frac{\max}{\min} \left\{
\begin{array}{l}
\displaystyle\sum_{i,j}\displaystyle\sum_m d_{i,j}(m)
\nu_{i,j}^{(m)} + \displaystyle\sum_{i,j}\displaystyle\sum_m
b_{i,j}(m) \mu_{i,j}^{(m)}\\
\\
\text{subject to}\\
\\
\bullet\quad\nu_{i,j}-x_0^j1_{i=0}=\displaystyle\sum_{k,l}
\tilde c_{k,l}(i,j)\mu_{k,l}, \  i+j \leq N,\ k+l\leq N\\
\quad\text{with}\quad \nu = \displaystyle\sum_m \nu^{(m)},\quad
\mu =
\displaystyle\sum_m \mu^{(m)}\\
\\
\bullet\quad \text{conditions \eqref{eq:H1}/\eqref{eq:H2} for
$\nu^{(m)}_{i,j}$, $\mu^{(m)}_{i,j}$,\quad $i+j\leq N$}
\end{array}
\right\}
\end{equation*}

In the case that the set $B$ is a half-line, the measures in
question will not have bounded support and as a consequence in the
above optimization problem the linear moment conditions
\eqref{eq:H1}/\eqref{eq:H2} are replaced by the quadratic moment
conditions described in Section \ref{sec:qmc}. The resulting
optimization problems are then semi-definite programming problems.
\cite{Lasser} provided convergence results for this SDP approach
in a diffusion setting for Asian, European and single barrier
options. Restricting to the case that $B$ is a finite interval we
show that the values $v^{(N)}_-, v^{(N)}_+$ of the linear programs
converge:

\begin{Prop}\label{prop:conv}
  Suppose that the system \eqref{eq:adftilde} has a unique solution and
  that $B$ is a finite interval. Then
  $$
v^{(N)}_- \uparrow v \quad\text{and}\quad v^{(N)}_+ \downarrow v.
  $$
  as $N\to\infty$.
\end{Prop}

{\bf Remark.} The presented approach can in principle be extended
to a multi-dimensional jump-diffusion $X$ with polynomial
coefficients. For example, if $B$ is a hyper-cube the adjoint
equations and the moment conditions take analogous forms. The
limitation in practice will be the capacity of the LP and SDP
solvers to deal with large size programs.

\section{Numerical examples}\label{sec:numerics}
For the numerical examples we have used Matlab and the LP solver lp\_solve.
The problems were set up in Matlab and then solved using the Matlab
interface to lp\_solve. The numerical outcomes were compared with Monte
Carlo results, implemented in Matlab using the Euler scheme.

We will illustrate the method by valuing four different options.

\subsection{A double knockout barrier option driven by the
Geometric Brownian motion.}

\begin{figure}[htbp]
\begin{center}
\includegraphics[width=0.75\textwidth]{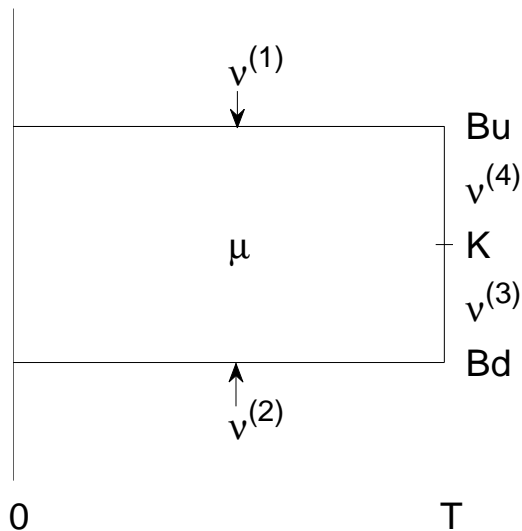}
\end{center}
\caption{Domain for the measures of double knockout option driven
by the Geometric Brownian motion.} \label{fig:BMdom}
\end{figure}

\noindent In this benchmark example we consider a European double
knock-out call option with underlying $S_t$ assumed to evolve as a
geometric Brownian motion. The value $v$ of such an option is
given by
\begin{align*}
v&=e^{-rT} E[(S_T-K)^+I_{\{\tau\geq T\}}]\quad\mathrm{where}\\
\tau&=\inf\{t\geq 0: S_t\notin [B_d,B_u]\}.
\end{align*}
For the ease of notation we will now drop the discounting
$e^{-rT}$. As a geometric Brownian motion has continuous paths,
we know that the time-space process $(t, S_t)$ will exit
$[0,T]\times[B_d, B_u]$ either if $S$ hits one of the barriers
$B_u$ or $B_d$ or maturity is reached, so that the support for the
exit location measure $\nu$ is
\begin{align*}
\Omega=\{[0,T]\times
\{B_d\}\}\cup\{[T]\times[B_d,B_u]\}\cup\{[0,T]\times\{B_u\}\}.
\end{align*}
The set $\Omega$ is partitioned into four parts with the
restricted measures
\begin{align*}
&\nu^{(1)}\textrm{ and } \nu^{(2)}\textrm{ with support on } [0,T]\\
&\nu^{(3)}\textrm{ and }\nu^{(4)}\textrm{ with support on }
[B_d,K] \textrm{ and } [K,B_u]\textrm{ respectively}.
\end{align*}
The expected occupation measure $\mu$ is supported on the domain
$[0,T]\times[B_d,B_u]$ -- See also Figure \ref{fig:BMdom} for an
illustration. 

Here the line-segments $[B_d,K]$ and $[K,B_u]$ were chosen in such
a way that the pay-off function restricted to each of those line
segments is zero or linear. Further note that the measures
$\nu^{(i)}$ can all be characterized by Hausdorff moment
conditions, since they are supported on line-segments. The value
of the option is then given by
\begin{align*}
v=\int_{\Omega}(x-K)^{+}I_{\{t=
T\}}\nu(dt,dx)=\nu^{(4)}_1-K\nu^{(4)}_0.
\end{align*}
Using the form \eqref{eq:genBGM} of the infinitesimal generator of
the Geometric Brownian motion, the basic adjoint equation for this
problem can be seen to be
\begin{align*}
&B_u^m\nu^{(1)}_n+B_d^m\nu^{(2)}_n+T^n\nu^{(3)}_m+T^n\nu^{(4)}_m-\\
&n\mu_{n-1,m}-\left(bm+\frac{\sigma^2}{2}m(m-1)\right)\mu_{n,m}=
t_0^nx_0^m.
\end{align*}
This is valid for all $n,m$ such that $n+m\leq N$, when we are
using all moments up to degree $N$. To complete the setup of the
problem we add the LP moment conditions for the measures
$\nu^{(i)}$ and $\mu$ with support as given above.

The numerical results for two given sets of parameter values are
given in Table \ref{tab:bmb015}. We can see that we get fast
convergence to the exact solution, which was calculated using the
formula from \cite{Pelsser}.

\begin{table}[t]
\centering {\small
\begin{tabular}{|lcccc|}
\hline
Case 1 & $b=0.1$ & $\sigma=0.1$ & &\\
\hline
Degree of moment &9& 10 & 11 & 12\\
Upper Bound & 0.9250 &   0.9211 &   0.9182  &  0.9161 \\
Relative Error & 1.61\% &   1.18\% & 0.86\% &   0.64\% \\
Lower Bound & 0.9096 &   0.9100 & 0.9102    &  0.9103 \\
Relative Error & 0.08\% &   0.03\% & 0.02\%  &   0.01\% \\
\hline
Exact solution & 0.9103 & &&\\
\hline
\hline
Case 2 & $b=0.2$ & $\sigma=0.2 $ &&\\
\hline
Degree of moment & 8 & 9 & 10 & 11 \\
Upper Bound   &  1.1656  & 1.1611 &   1.1569 &   1.1534\\
Relative Error  &  2.06\% &  1.66\%  &    1.29\%  &  0.99\%\\
Lower Bound   &  1.1064  & 1.1163 &   1.1256 &   1.1293\\
Relative Error  &  3.13\% & 2.26\%  &   1.45\%  &  1.13\%\\
\hline
Exact solution & 1.1421  &&&\\
\hline
\end{tabular}}
\caption{Numerical results for the Double knockout Barrier option
with the underlying modelled by the Geometric Brownian motion. The
option parameters are $B_u=5$, $B_d=1$, $K=1.3$, $x_0=2$, $t_0=0$
and $T=1$.}\label{tab:bmb015}
\end{table}

\subsection{A double knockout barrier option driven by the Variance gamma
process.} 
\begin{figure}[t]
\begin{center}
\includegraphics[width=0.75\textwidth]{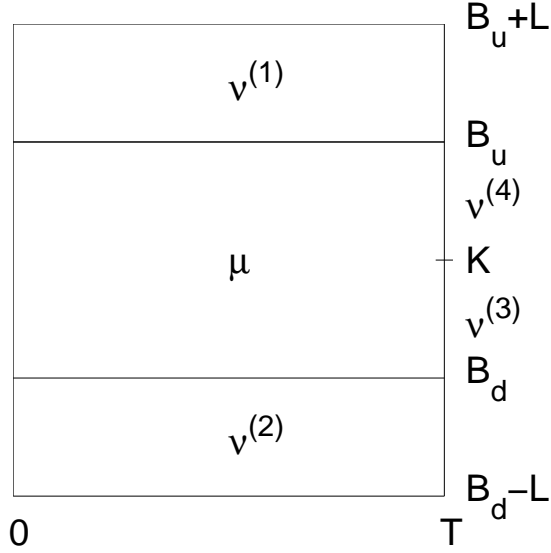}
\end{center}
\caption{Domain for the truncated measures of a double knockout option in the
Variance Gamma case.}
\label{fig:levydom}
\end{figure}

In this example we consider again a double knock-out
option but now driven by a Variance Gamma process, which as
described in Section \ref{sec:prob}. Since the Variance Gamma
process is a finite activity jump process, it will not hit the
barrier but jump across it. As a consequence the exit location
measure $\nu$ is supported on
$$\Omega = \{[0,T]\times[B_u,\infty)\}\cup
\{\{T\}\times[B_d,B_u]\}\cup \{[0,T]\times (-\infty,B_d]\}.$$ In
order to be able to calculate the value $v$ of the option using
the LP moment conditions, we will adjust the L\'evy measure $\eta$
as described in Section \ref{sec:adjoint}, to achieve bounded
support. In this case we observe that any jump with absolute size
larger than $L = B_u-B_d$ will trigger an immediate knock-out. We
note that the probability that no such a jump occurs before
maturity is $p_*= e^{-\lambda_* T}$ where $\lambda_*=
\eta(\R\backslash [-L,L])$ with $\eta$ the Variance Gamma L\'{e}vy
measure given in \eqref{eq:VGeta}. We thus truncate the L\'{e}vy
measure $\eta$ by restricting it to absolute jump-sizes smaller
$L$:
$$\tilde\eta(dx) = I_{\{|x|<L\}}\eta(dx). $$
The value of the option in terms of moments of these measures is
then
\begin{align*}
v= p_* \int_{\Omega}(x-K)^{+}I_{\{t=T\}}\nu(dt,dx)=
p_*[\nu^{(4)}_1-K\nu^{(4)}_0].
\end{align*}
where the support of the truncated exit location measure is
\begin{align*}
\tilde\Omega=\{[0,T]\times[B_u,B_u+L]\}\cup
\{[T]\times[B_d,B_u]\}\cup \{[0,T]\times[B_d-L,B_d]\}
\end{align*}
and the four restrictions of $\nu$ are
\begin{align*}
&\nu^{(1)}\quad\textrm{supported on } [0,T]\times[B_u,B_u+L]\\
&\nu^{(2)}\quad\textrm{supported on } [0,T]\times[B_d-L,B_d]\\
&\nu^{(3)}\ \text{and}\ \nu^{(4)}\quad\textrm{supported on }
[B_d,K] \ \text{and}\ [K,B_u]
\end{align*}
The domain of the truncated measures is shown in Figure
\ref{fig:levydom}.
Denoting by
\begin{align}\label{eq:ck}
c(k)=\int_{-L}^{L} y^kk(y)dy
\end{align}
the moments of the truncated L\'evy measure $\tilde\eta$ and
taking note of the form \eqref{eq:genVG} the infinitesimal
generator, we find the basic adjoint equation
\begin{align*}
\nu^{(1)}_{n,m}+\nu^{(2)}_{n,m}+T^n\left(\nu^{(3)}_m+\nu^{(4)}_m\right)-
n\mu_{n-1,m}-bm\mu_{n,m-1}\\
- \sum_{k=1}^m\binom{m}{k}c(k)\mu_{n,m-k}=0^nx_0^m
\end{align*}
for all $n,m$ such that $m+n\leq N$. As before, to complete the LP
problem we add the appropriate LP moment conditions for the
measures $\nu^{(i)}$ and $\mu$ with support as given above.

Numerical results can be found in the Table \ref{tab:VG}. The
relative error was calculated using the arithmetic mean of the
upper and lower bounds and the Monte Carlo outcome (taking the
latter as the `true' result). Studying the results we see that we
get tight bounds within 8 or 9 moments. Beyond 10 or 11 moments we
experienced instabilities with the LP solver.

In Case 1 the execution times are shown, which should be
compared to the execution time for the Monte Carlo simulation that
was around 52 minutes. As we employed a basic Euler scheme for the
Monte Carlo simulation the speed of convergence of the Monte Carlo
simulation could be improved by using more specialised Monte Carlo
schemes and also by changing to a compiling programming language.
However considering the time difference, it should be clear that
the method of moments will still be considerably faster.


\begin{table}[t]
\centering {\small
\begin{tabular}{|lcccc|}
\hline
Case 1 & $G=8$& $M=12$ & $p_*=1.0000$&\\
\hline
Degree of moment & 7 & 8 & 9 & 10 \\
Upper Bound & 0.5045 & 0.5030 & 0.5022 & 0.5017\\
Lower Bound & 0.4946 & 0.4983 & 0.4987 & 0.4994\\
Relative Error & 0.13\% & 0.09\% & 0.05\% & 0.07\%\\
Cpu Time & 1.262s & 3.245s & 6.236s & 17.936s\\
\hline
Monte Carlo &0.5002 & Std Error & 0.0005 &\\
\hline
\hline
Case 2 & $G=4$& $M=10$ & $p_*=1.0000$ &\\
\hline
Degree of moment & 6 & 7 & 8 & 9 \\
Upper Bound & 0.5158 & 0.5151 & 0.5135 & 0.5115\\
Lower Bound & 0.4857 & 0.4886 & 0.4943 & 0.4958\\
Relative Error & 0.39\% & 0.17\% & 0.24\% & 0.19\%\\
\hline
Monte Carlo & 0.5027 & Std Error &0.0008 &\\
\hline
\hline
Case 3 & $G=8$& $M=8$ &$p_*=1.0000$&\\
\hline
Degree of moment & 5 & 6 & 7 & 8 \\
Upper Bound &  0.5133  &  0.5078  &  0.5049 &   0.5033\\
Lower Bound & 0.4682 &   0.4894  &  0.4917 &   0.4957\\
Relative Error &1.71\% & 0.14\% & 0.20\% & 0.04\%\\
\hline
Monte Carlo & 0.4993 & Std Error& 0.0006 &\\
\hline
\hline
Case 4 & $G=3$& $M=6$ &$p_*=0.9998$&\\
\hline
Degree of moment  & 6 & 7 & 8 & 9 \\
Upper Bound & 0.5277  &  0.5237  &  0.5197 &   0.5182\\
Lower Bound & 0.4672  &  0.4720  &  0.4745 &   0.4772\\
Relative Error &1.32\% & 1.24\% & 1.39\% & 1.27\%\\
\hline
Monte Carlo & 0.5041 & Std Error & 0.0011 &\\
\hline
\end{tabular}}
\caption{Numerical results for the Double knockout barrier option
driven by the Variance Gamma process. The option parameters are in
all cases, $B_u=1$, $B_d=-1$ and $K=-0.3$, the model parameters
are $b=0.2$, $x_0=0$, $T=1$ and $C=0.5$. The truncation size is
$L=2$ in all cases, and $p_*$ is the probability that no jump of absolute size
larger than 2 occurs before $T$. 
}\label{tab:VG}
\end{table}

\subsection{The American Corridor under a CIR model with constant interest
rate} An American corridor is a contract traded in the Foreign
exchange markets that pays a continuous rate until either the
underlying leaves the corridor or maturity is reached, whichever
comes earlier (see for example \cite{Weithers} or \cite{Wystup}
for background). The value of this contract is given by
\begin{align*}\label{eq:ampay}
v&=E\left[\int_0^\tau
e^{-rt}dt\right]=\int_\Theta\mu(dt,ds)=
\mu_{0,0}\quad\textrm{where}\\
\tau&=T\land\inf\{t\geq 0:S_t\notin[B_d,B_u]\},\nonumber
\end{align*}
with $\Theta=\{[0,T]\times[B_d,B_u]\}$. We will model the
underlying as a Cox Ingersoll Ross (CIR) process, evolving
according to the SDE \eqref{eq:SDECIR}. Since the CIR model is
continuous the supports of the different measures are given as
follows:
\begin{align*}
&\nu^{(1)}\textrm{ and } \nu^{(2)}\textrm{ supported on } [0,T]\\
&\nu^{(3)}\textrm{ supported on } [B_d,B_u]\\
&\mu\textrm{ is supported on } [0,T]\times[B_d,B_u]
\end{align*}
In view of the form of the infinitesimal generator for the CIR
process \eqref{eq:genCIR} we can now assemble the basic adjoint
equation for this problem,
\begin{align*}
&B_u^m\nu^{(1)}_n+B_d^m\nu^{(2)}_n+T^n\nu^{(3)}_m-n\mu_{n-1,m}+\\
&(am+r)\mu_{n,m}-\left(abm+\frac{\sigma^2}{2}m(m-1)\right)\mu_{n,m-1}=
t_0^nx_0^m
\end{align*}
for all $m,n$ such that $m+n\leq N$, and add appropriate LP moment
conditions as before.

The results are reported in Table \ref{tab:Amcorr}. We observe
that tight bounds are achieved, and that in cases 1 and 2 the upper
bounds are accurate for 9-10 moments, with relative errors $0.16\%$
and $0.12\%$. We also see that the speed of
convergence varies with the particular parameter values.

\begin{table}[t]
\centering {\small
\begin{tabular}{|lcccc|}
\hline
Case 1 & $\sigma=0.2$& $r=0.1$ &&\\
\hline
Degree of moment & 10& 11 & 12 & 13 \\
Upper Bound & 0.9516  &  0.9516  &  0.9516 &   0.9516\\
Lower Bound & 0.9274  &  0.9345  &  0.9391 &   0.9421\\
Relative Error &1.12\% & 0.74\% & 0.50\% & 0.34\%\\
\hline
Monte Carlo & 0.9501 & Std Error & 0.0002 &\\
\hline
\hline
Case 2 & $\sigma=0.2$& $r=0.05$ &&\\
\hline
Degree of moment & 9 & 10 & 11 & 12 \\
Upper Bound & 0.9754  &  0.9754  &  0.9754 &   0.9754\\
Lower Bound & 0.9394  &  0.9504  &  0.9577 &   0.9624\\
Relative Error &1.73\% & 1.16\% & 0.79\% & 0.54\%\\
\hline
Monte Carlo & 0.9742 & Std Error & 0.0002 &\\
\hline
\hline
Case 3 & $\sigma=0.3$& $r=0.1$ &&\\
\hline
Degree of moment & 11 & 12 & 13 & 14\\
Upper Bound & 0.9343  &  0.9325  &  0.9315 &   0.9307\\
Lower Bound & 0.8961  &  0.9024  &  0.9067 &   0.9095\\
Relative Error &0.76\% & 0.52\% & 0.34\% & 0.29\%\\
\hline
Monte Carlo & 0.9222 & Std Error & 0.0011 &\\
\hline
\end{tabular}}
\caption{Numerical results for the American Corridor modelled by
the Cox Ingersoll Ross model. The problem parameters are
$B_d=0.5$, $B_u=1.5$, $x_0=1$ and $T=1$, for the model parameters
$a=0.5$ and $b=1$ are fixed.} \label{tab:Amcorr}
\end{table}

\subsection{Double No Touch option under the exponential Variance Gamma
process with a non constant interest rate} A double no touch
option pays one unit at maturity if the underlying has not crossed
either of the barriers $B_d$ or $B_u$. Its value can be expressed
as
\begin{align*}
v&=E[e^{-\alpha_T}I_{\{\tau\geq T\}}]\quad\textrm{where}\\
\tau&=\inf\{t\geq 0:S_t\notin[B_d,B_u]\}.
\end{align*}
Letting the underlying $S_t$ be an exponential Variance Gamma
process we note that the stopping time $\tau$ is equivalent to
\begin{align*}
\tau=\inf\{t\geq 0:X_t\notin[\log(B_d),\log(B_u)]\}
\end{align*}
where $X_t$ is a Variance Gamma process. Under no-arbitrage
pricing the process $e^{-\alpha_t}S_t = e^{-\alpha_t + X_t}$ needs
to be a martingale which is equivalent to the requirement that
\begin{align*}
b(t)+\int_{-\infty}^{\infty}(e^x-1)\eta(dx)=r(t)
\end{align*}
so that $b(t)$ is determined by our choice of $r(t)$. For the
Variance Gamma process
\begin{align*}
c=\int_{-\infty}^{\infty}(e^x-1)\eta(dx)=C\left(\log\left(\frac{G}{1+G}\right)+
\log\left(\frac{M}{1-M}\right)\right)
\end{align*}
With these points in mind we find,
\begin{align*}
(A-r(t))f(t,x)&=\frac{\partial f}{\partial t}+b(t)\frac{\partial f}{\partial x}
+\int[f(t,x+y)-f(t,x)]\eta(dx)-r(t)f(t,x)\\
&=\frac{\partial f}{\partial t}+(r(t)-c)\frac{\partial f}{\partial x}
+\int[f(t,x+y)-f(t,x)]\eta(dx)-r(t)f(t,x)
\end{align*}
We have chosen to study interest rates of the type
\begin{align*}
r(t)=r_b+r_st^2.
\end{align*}
Since there is no need too split the exit location measure at
maturity, the domain  of $\nu$ only needs to be split into three
parts $K_1=[0,T]\times[0,B_d]$, $K_2=[0,T]\times[B_u,\infty)$ and
$K_3=\{T\}\times[B_d,B_u]$, so that in this case the value $v$ of
the option can be expressed in terms of moments as follows
\begin{align*}
v&=E[e^{-\alpha_T}I(\tau\geq T)]=\int_\Omega I(t= T)\nu(dt,ds)\\
&=\int_{K_3}\nu^{(3)}(ds)=\nu^{(3)}_0
\end{align*}
where as before $\nu^{(i)} =\nu(\cdot \cap K_i)$. In terms of
moments the basic adjoint equation is then given by
\begin{align*}
\nu^{(1)}_{n,m}+\nu^{(2)}_{n,m}+T^n\nu^{(3)}_m-t_0^nx_0^m=
n\mu_{n-1,m}-r_b\mu_{n,m}-r_s\mu_{n+2,m}\\+(r_b-c)m\mu_{n,m-1}+
r_sm\mu_{n+2,m-1} +\sum_{k=1}^m\binom{m}{k}c(k)\mu_{n,m-k},
\end{align*}
where $c(k)$ is given in \eqref{eq:ck}.

The numerical results are presented in Table \ref{tab:dnt}. In all
cases we see tight bounds nicely agreeing with the Monte Carlo
simulation result. We also observe that the upper bound is very
accurate already for a small number of moments. For example, for 7
moments, the relative errors of the upper bounds in the three
different cases are 0.032\%, 0.076\% and 0.069\%, respectively.

\begin{table}[t]
\centering {\small
\begin{tabular}{|lcccc|}
\hline
Case 1 & $r_b=0.05$& $r_s=0.05$ &&\\
\hline
Degree of moment & 6 & 7 & 8 & 9 \\
Upper Bound & 0.9356 & 0.9355 & 0.9355 & 0.9355 \\
Lower Bound & 0.8453 & 0.8757 & 0.9042 & 0.9143\\
Relative Error & 4.79\% & 3.17\% & 1.64\% & 1.10\%\\
\hline
Monte Carlo &0.9352 & Std Error & 0.0002 &\\
\hline
\hline
Case 2 & $r_b=0.05$& $r_s=0.1$ &&\\
\hline
Degree of moment & 6 & 7 & 8 & 9 \\
Upper Bound & 0.9203 & 0.9201 & 0.9200 & 0.9200 \\
Lower Bound & 0.8196 & 0.8533 & 0.8836 & 0.8957\\
Relative Error & 5.38\% & 3.56\% & 1.91\% & 1.26\%\\
\hline
Monte Carlo  &0.9194 & Std Error & 0.0002 &\\
\hline
\hline
Case 3 & $r_b=0.1$& $r_s=0.1$ &&\\
\hline
Degree of moment & 7 & 8 & 9 & 10\\
Upper Bound & 0.8752 & 0.8752 & 0.8752 & 0.8752 \\
Lower Bound & 0.7980 & 0.8319 & 0.8449 & 0.8565\\
Relative Error & 6.68\% & 4.73\% & 2.84\% &
2.09\%\\
\hline
Monte Carlo &0.8746 & Std Error & 0.0002 &\\
\hline
\end{tabular}}
\caption{Numerical results for the double no touch option with
barriers at $B_u=2$ and $B_d=0.5$ and maturity $T=1$, driven by an
exponential Variance Gamma process with parameters $C=0.5$, $G=8$
and $M=12$, and with $S_0=1$. Note that the upper bounds are very
accurate already for 6-7 moments, with relative errors less than
0.1\%.}\label{tab:dnt}
\end{table}

\section{Conclusion}
We have presented a method of moments approach that can be used to
price double barrier-type options driven by `polynomial'
jump-diffusions, allowing for a non-constant (deterministic or
stochastic) interest rate. An infinite-dimensional linear program
was derived, which was then approximated, depending on the choice
of moment conditions, either by a sequence of LP problems, or by a
sequence of SDP problems. Although the SDP-type problems may be
theoretically more appealing as the SDP method naturally handles
measures with unbounded support, further development of stable SDP
solvers would be needed for this method to be truly usable in
practice. Since, on the other hand, the LP solvers are in a more
advanced state of development and several (commercial) LP solvers
are available capable of solving (large scale) LP problems, we
focussed on the LP approach. We formulated the approximating
programs as LP problems by using truncation, and provided
theoretical convergence results for this approach. We illustrated
the method with numerical examples, using the Matlab interface of
the solver {\tt lp\_solve}, and compared the outcomes with Monte
Carlo simulation
results. We found that accurate results with tight upper and lower
bounds were obtained with a small number of moments in most of the
examples, and observed that the algorithm was significantly faster
than Monte Carlo simulation.

\newpage
\appendix
\section{Proofs}

\subsection{Proof of Proposition \ref{prop:conv}}
Since the number of equations grows with $N$, it
follows that $v^{(N)}_-$ is monotone increasing, since the minimum
taken over a smaller set of elements is larger.

Note that $v$ is a finite linear combination of moments. Because
of the fact that the support of the different measures is compact,
if follows that each moment is bounded, so that $v$ is bounded and
$v^{(N)}_-$ is the minimization of a linear function over a
bounded set. Thus, the minimum $v^{(N)}_-$ is finite and attained
at a vector $q^N=(q^N_i)_i$ that satisfies the corresponding
linear system of equations. Thus, for each fixed $i$ there exists
a $q^*_i$ such that, for $N$ along a subsequence, $q^N_i \to
q^*_i$. In fact, by a diagonal argument it follows that there
exists a subsequence $\tilde N$ such that, as $\tilde N\to\infty$,
$$q^{\tilde N}_i \to q^*_i\quad\text{for all }i.$$
Clearly, $q^*$ satisfies the infinite system and thus under the
assumption that there exists a unique sequence that solves the
infinite system, it follows that $q^*$ must be equal to
$(\mu_i,\nu_i)_i$. Moreover, $\mu$ and $\nu$ are the unique
measures corresponding to these moments, as they are both
moment-determinate. Thus, $v = L(q^*)$ and $v^{(N)}_- \uparrow
L(q^*)$

The proof of the convergence of the sequence $(v^{(N)}_+)$ is
similar and omitted.\exit

\subsection{Proof of Proposition \ref{prop:dynkin}}

We will show the following lemma:

\begin{Lemma}\label{lem:dynkin} For any bounded stopping time $\tau$ and $f\in
C^{1,2}$ with
\begin{equation}\label{eq:cint}
E\left[\int_0^\tau e^{-\alpha_t}\left[\sigma^2\frac{\partial
f}{\partial x}\right]^2(t,X_t)\right] + E\left[ \int_0^\tau
e^{-\alpha_t} |g(t,X_t,y)|\Lambda(dy)dt\right] < +\infty,
\end{equation}
with
$$
g(t,x,y) = f(t,x + y\lambda(t,x)) - f(t,x) - \frac{\partial
f}{\partial x}(t,x)\lambda(t,x)y1_{|y|< 1},
$$
eqn. \eqref{eq:intbae} holds true.
\end{Lemma}

The proposition is a direct consequence of this lemma, since under
the condition \eqref{eq:int} the integrability conditions are
satisfied for each monomial $t^ix^j$.

{\it Proof of Lemma \ref{lem:dynkin}:} Applying (a general form
of) It\^{o}'s lemma to the stochastic process
$e^{-\alpha_t}f(t,X_t)$ (which is justified as $f\in C^{1,2}$)
shows that
\begin{eqnarray}\label{eq:ito}
e^{-\alpha_t}f(t,X_t) - f(0,X_0) &=& M_t + \int_0^t
e^{-\alpha_s}(Af - r f)(s, X_s)ds
\end{eqnarray}
where $Af$ is given in \eqref{A} and $M_t$ is the local martingale
given by
$$
M_t = \int_0^te^{-\alpha_t}\sigma(t,X_t)dW_t +
\int_{[0,t]\times\R} e^{-\alpha_t} g(t,X_t,y)\phi(dy,dt),
$$
where $\phi$ denotes the compensated jump measure associated to
$J$ (with compensator $\Lambda(dy)dt$). Note that the identity
\eqref{eq:ito} remains valid with $t$ replaced by $t\wedge \tau$
with $M^\tau=\{M_{t\wedge\tau}, t\ge 0\}$ a local martingale.
Under the integrability conditions \eqref{eq:cint} it holds that
that $M^\tau$ is a zero mean martingale. Taking expectations in
\eqref{eq:ito} shows thus that
$$
E[e^{-\alpha_t}f(t,X_t)] - f(0,x) = E\left[\int_0^t
e^{-\alpha_s}(Af - r f)(s, X_s)ds\right]
$$
which shows that \eqref{eq:intbae} is valid. \exit

\subsection{Proof of Corollary \ref{cor:dynkin}}

Since the jump-diffusion $\tilde X$ with drift $b$, volatility
$\sigma$, L\'{e}vy measure $\tilde \Lambda$ and discounting
$\tilde r$ satisfies \eqref{eq:int}, Proposition \ref{prop:dynkin}
yields that the measures $\tilde \nu$ and $\tilde \mu$ corresponding
to $\tilde X$ satisfy
$$
\tilde\nu_{i,j}-x_0^j1_{i=0}=\sum_{k,l}\tilde
c_{k,l}(i,j)\tilde\mu_{k,l}\quad i,j=0,1,2,\ldots.
$$
where
$$(\tilde Af_{ij}-\tilde rf_{ij})(t,x)=\sum_{k,l}\tilde
c_{k,l}(i,j)t^lx^k,$$ with $f_{ij}=t^ix^j$. To complete the proof
we will now show that
\begin{equation}\label{A.v}
v = \int h(t,x) \tilde\nu(dt,dx) + \int g(t,x)\tilde\mu(dt,dx).
\end{equation}
Denoting by $\rho$ the first time that a jump of $J$ of size
smaller than $L_-$ or larger than $L_+$ and let
$$\tilde\tau=\inf\{t\ge0: \tilde X_t\notin B\}.$$ Then, if
$\rho>\tau$ it holds that $X_{t\wedge\tau} = \tilde
X_{t\wedge\tau}$ for all $t\ge 0$ and in particular
$$
\tau =\tilde\tau \quad\text{and}\quad X_\tau = \tilde X_\tau =
\tilde X_{\tilde\tau}.
$$
Also, since it is assumed that $h(t,x)=0$ for $x\notin B$, we have
that $h(\tau, X_\tau) = h(\tau, X_\tau)1_{\{\rho > \tau\}}$.
Taking note of these observations, it follows that
\begin{eqnarray*}
E[e^{-\alpha_\tau}h(\tau, X_\tau)] &=& E[e^{-\alpha_\tau}h(\tau,
X_\tau)1_{\{\rho > \tau\}}]\\
&=& E[e^{-\alpha_{\tilde\tau}}h(\tilde\tau,
X_{\tilde\tau})1_{\{\rho > \tilde\tau\}}]\\
&=&
E[e^{-\alpha_{\tilde\tau}-\tilde\lambda\tilde\tau}h(\tilde\tau,
X_{\tilde\tau})] \\
&=& E[e^{-\tilde\alpha_{\tilde\tau}}h(\tilde\tau, X_{\tilde\tau})]
\end{eqnarray*}
where $\tilde\lambda := \Lambda(\R\backslash[L_-, L_+])$ and we
used that $\rho$ follows an exponential distribution with mean
$\tilde\lambda^{-1}$, independent of $\tilde X$. Similarly,
\begin{eqnarray*}
E\left[\int_0^\tau e^{\alpha_s}g(s, X_{s-})ds\right] &=&
E\left[\int_0^\tau
e^{-\alpha_s}g(s, X_{s-})1_{\{s\leq\rho\}}ds\right]\\
&=& E\left[\int_0^{\tilde\tau} e^{-\alpha_s}g(s, \tilde
X_{s-})1_{\{s\leq\rho\}}ds\right]\\
&=& E\left[\int_0^{\tilde\tau} e^{-\alpha_s - \tilde\lambda s}g(s,
\tilde
X_{s-})ds\right]\\
&=& E\left[\int_0^{\tilde\tau} e^{-\tilde\alpha_s}g(s, \tilde
X_{s-})ds\right].
\end{eqnarray*}
The two identities imply that \eqref{A.v} holds true, and the
proof is complete.\exit

\section{Semi-definite moment conditions}

The moment matrices are defined as follows:

\paragraph{Moment Matrices}
Let
\begin{align}\label{eq:base}
(x^\alpha,|\alpha|\leq
k)=(1,x_1,\ldots,x_n,x_1^2,x_1x_2,\ldots,x_1^k,
x_1^{k-1}x_2,\ldots,x_n^k),
\end{align}
be the usual basis of polynomials in $n$ variables with degree at
most $k$.

Given a series of moments of a measure
$m=\{m_\alpha,\alpha\in\N^n\}$ let $\hat{m}=\{\hat{m}_i,i\in\N\}$
be that sequence ordered in accordance with \eqref{eq:base}. The
Moments matrix $M_k(m)$ is then defined as
\begin{align*}
M_k(m)(1,i)=M_k(m)(i,1)=\hat{m}_{i-1}, \textrm{for } i=1,\ldots,k+1,\\
M_k(m)(1,j)=m_\alpha \textrm{ and } M_k(m)(i,1)=m_\beta\Rightarrow
M_k(m)= m_{\alpha+\beta}
\end{align*}
where $M_k(m)(i,j)$ is the $(i,j)$-entry of the matrix $M_k(m)$.
\paragraph{Localising Matrices}
Given a polynomial $q$ with coefficients ($q_\alpha$) in the basis
\eqref{eq:base}. If $\beta(i,j)$ is the $\beta$ subscript of the
$(i,j)$-entry of the moment matrix $M_k(m)$ then the localising
matrix is defined by
\begin{align*}
M_k(q,m)(i,j)=\sum_\alpha q_\alpha m_{\beta(i,j)+\alpha}
\end{align*}
The choice of the function $q$ depends on the support of the
measure. For example, in the one-dimensional setting we have three
cases,
\begin{enumerate}
\item Support on $[a,b]$, with function $q=(b-x)(x-a)$ \item
Support on $[a,\infty)$, with function $q=(x-a)$ \item Support on
$(-\infty,a]$, with function $q=(a-x)$
\end{enumerate}
In terms of moment and localizing matrices the characterization is
then as follows: Given $m=(m_0,m_1,\ldots,m_{2r})$ the condition
that $M_r(m)$ and $M_{r-1}(q,m)$ are positive semi-definite are
sufficient conditions for the elements of $m$ to be the first
$2r+1$ moments of a measure supported on the appropriate interval.


\newpage

\begin{thebibliography}{10}

\bibitem{BoyLev}
S.~I. Boyarchenko and S.~Levendorskii.
\newblock {\em {Non-Gaussian Merton-Black-Scholes Theory}}.
\newblock World Scientific Publishing Co Pte Ltd, 2002.

\bibitem{CarrCrosby}
P.~Carr and J~Crosby.
\newblock {A class of L\'evy process models with almost exact calibration to
  both barrier and vanilla fx options}.
\newblock Working paper available September 2008 at
  \url{http://www.john-crosby.co.uk/pdfs/Carr_Crosby_DNT&Vanilla_Levy_PDF.pdf},
  2008.

\bibitem{ContTankov}
R.~Cont and P.~Tankov.
\newblock {\em {Financial Modelling With Jump Processes}}.
\newblock Chapman \& Hall/CRC, 2004.

\bibitem{Crosby}
J.~Crosby.
\newblock {A multi-factor jump-diffusion model for commodities}.
\newblock {\em Quantitative Finance}, 8(2):181--200, 2008.

\bibitem{Cuchiero}
C.~Cuchiero, D.~Filipovi\'c, and J.~Teichmann.
\newblock {Affine Models}.
\newblock Working paper available September 2008 at
  \url{http://www.vif.ac.at/filipovic/PAPERS/vif8.pdf}, 2008.

\bibitem{CurtoFial}
R.E Curto and L.A. Fialkow.
\newblock {Recursivness, Positivity and Truncated Moment Problems}.
\newblock {\em Houston J. Math.}, 17(4), 1992.

\bibitem{Linetsky}
D.~Davydov and V.~Linetsky.
\newblock {Pricing Options on Scalar Diffusions: An Eigenfunction Expansion
  Approach}.
\newblock {\em Operations Research}, 51(2), 2003.

\bibitem{Feller}
W.~Feller.
\newblock {\em {An Introduction to Probability Theory and Its Applications}},
  volume~2.
\newblock Wiley, 2 edition, 1971.

\bibitem{Geman}
H.~Geman.
\newblock {\em {Commodities and commodity derivatives}}.
\newblock Wiley, 2005.

\bibitem{GeYor}
H.~Geman and M.~Yor.
\newblock {Pricing and Hedging Double-Barrier Options: A Probabilistic
  Approach}.
\newblock {\em Mathematical Finance}, 6(4), 1996.

\bibitem{Stockbridge}
K.~Helmes, S.~R{\"{o}}hl, and R.H. Stockbridge.
\newblock {Computing Moments of the Exit Time Distribution for Markov Processes
  by Linear Programming}.
\newblock {\em Operations Research}, 49(4), 2001.

\bibitem{KreinNudel}
M.~Krein and A.~Nudel'man.
\newblock {The Markov Moment Problem and Extremal Problems}.
\newblock In {\em Transl. Math. Monograps}, volume~50. American Mathematical
  Society, 1977.

\bibitem{Lasser}
J.B. Lasserre, T.~Prieto-Rumeau, and M.~Zervos.
\newblock {Pricing a class of Exotic Options via Moments and SDP Relaxations}.
\newblock {\em Mathematical Finance}, 16(3), 2006.

\bibitem{MCC}
D.B. Madan, P.~Carr, and E.~Chang.
\newblock The {V}ariance {G}amma process and option pricing model.
\newblock {\em Eur. Finance Rev.}, 2:79--105, 1998.

\bibitem{Pelsser}
A.~Pelsser.
\newblock {Pricing double barrier options using Laplace transforms}.
\newblock {\em Finance and Stochastics}, 4(1), 2000.

\bibitem{Schoutens}
W.~Schoutens.
\newblock {\em {L\'evy Processes in Finance: Pricing Financial Derivatives}}.
\newblock Wiley, 2003.

\bibitem{Sepp}
A.~Sepp.
\newblock {Analytical Pricing of Double-Barrier Options Under A
  Double-Exponential Jump Diffusion Process: Applications of Laplace
  Transform}.
\newblock {\em International Journal of Theoretical and Applied Finance}, 7(2),
  2004.

\bibitem{ShohatTamarkin}
J.~A. Shohat and J.~D. Tamarkin.
\newblock {\em {The Problem of Moments}}.
\newblock American Mathematical Society, 1943.

\bibitem{Stoy}
J.~Stoyanov.
\newblock {Moment Problems Related to the Solutions of Stochastic Differential
  Equations}.
\newblock In {\em Stochastic Theory and Control: Proceedings of a Workshop held
  in Lawrence, Kansas}, volume 280, pages 459--469. Springer, 2002.

\bibitem{Weithers}
T.~Weithers.
\newblock {\em {Foreign Exchange: A Practical Guide to the FX Markets}}.
\newblock Wiley, 2006.

\bibitem{Wystup}
U.~Wystup.
\newblock {\em {FX Options and Structured Products}}.
\newblock Wiley, 2007.

\end{thebibliography}
\end{document}